\documentclass[preprint,authoryear,12pt]{elsarticle}
\usepackage{amssymb}
\journal{New Astronomy}

\begin{document}

\begin{frontmatter}

\title{Identification of Stellar Sequences in various Stellar Systems : ESO65-SC03, TEUTSCH 106, TURNER 6}
\author[1]{Gireesh C. Joshi}
\ead{gchandra.2012@rediffmail.com}
\address[1]{Department of Physics, Kumaun University Uttarakhand (INDIA)-263001}

\begin{abstract}
The spatial morphological study of studied clusters is carried out through the identified probable members within them. The field stars decontamination is performed by the statistical cleaning approach (depends on the magnitude and colour of stars within the field and cluster regions). The CMRD (colour magnitude ratio diagram) approach is used to separate stellar sequences of the cluster systems. The age, distance and reddening of each cluster are estimated through the visual inspection of best fitted isochrone in colour magnitude diagrams(CMDs). The mean proper motion values of clusters are obtained through the extracted data from PPMXL and UCAC4 catalogs. Moreover, these values are varying according to the extracted data-set from these catalogues. This variation is occurred due to their different estimation efficiency of proper motions. The TCR (two colour ratio) and TCMR (two colour magnitude ratio) values of each cluster are determined by utilizing the WISE and PPMXL catalogues, these values are found abnormal for TEUTSCH 106. In addition, the TCMR values are similar to TCR values at longer wavelength, whereas both values are far away to each other at shorter wavelength. The fraction of young stellar objects (YSOs) is also computed for each cluster.
\end{abstract}


\end{frontmatter}


\section{Introduction}
\label{sec:intro}
The open star clusters (OSCs) are distributed in the galactic plane of spiral and irregular galaxies. These objects are used for the stellar and galactic investigations [\citet{ha05} and references within it]. Since, OSCs are formed through the collapse and fragmentation of a turbulent molecular cloud \citep{hap,bat}, therefore, they become ready samples to study of the stellar evolution history. OSCs is also used to constraint the stellar formation processes in the Galactic disk \citep{di12}. Such studies are easily carry out by CMDs due to stars, having the same distance, of the same chemical composition and age; but they having different mass \citep{fri,has}. OSCs are dynamic and loosely gravitational bound systems \citep{jr15}. Generally, these objects are identified due to their stellar enhanced compare to the nearby Sky. They are also disrupted at the time of revolving around the Galactic center through the close encounters with another neighbor cluster and clouds of gas. Thus, the stellar density of any cluster decreases with the time due to the dispersion and internal close encounters of its members. The OSCs can be divided into the dense and sparse OSCs according to their stellar density. The main sequence (MS) of sparse one is easily identified compare to dense ones. Moreover, the estimation  accuracy of parameters of OSC depends on the identified MS. Thus, each cluster shows unique stellar sequence (i.e. MS) of dynamical stars. There are several examples of the CMDs of clusters [such as, NGC 4755 \citep{bo06}, M11 \citep{sa05} etc.], having apparent stellar sequences more than one. One of these sequences is referred as the MS of studied cluster, whereas other may be the Sub- or red-giant branch. The cluster's parameter dependency on stellar sequence may be understood through the comparative study of some sparse and dense clusters. \cite{jo16} has been proposed the colour-magnitude ratio diagram method (CMRD) for separating the stellar sequences and stellar membership of clusters. In this connection, we are follow up the CMRD method for the membership estimation of clusters. Here, we are presented the spatial morphological  study of one cluster remnant [ESO 65 03 \citep{jo15}] and one dense cluster regions (TURNER 6 + TEUTSCH 106). These clusters are listed on SIMBAD\footnote{http://simbad.u-strasbg.fr/guide/index.htx} website, which has provided the clusters database of studied and poorly studied clusters. The data-sets of present studied clusters have been extracted from the PPMXL \citep{ros}, UCAC4  \citep{zac} and WISE \citep{wri} catalogues.\\
Present manuscript is organized as follows. The previous studies of clusters are described in Section~\ref{pr02}. The existence of clusters is discussed in Section \ref{ce03}. The parametric results are estimated in Section \ref{pa04}. Sections \ref{tc05} and \ref{yo06} are devoted to two colours-magnitude ratio diagrams and identification of young stellar objects, respectively. The final conclusions and discussion are stated in Section \ref{co07}. 
\section{Previous studies}
\label{pr02}
In the present sample of clusters, we have taken one cluster remnant (ESO65SC03 or ESO 65 03) and one dense cluster regions (TURNER 6). The selected field region of TURNER 6 is also covered the field region of TEUTSCH 106. These stellar systems are situated at the southern hemisphere of the Sky. \cite{dias1, dias2} and \cite{khar} have been compiled the catalogues for the optically visible clusters and candidates. The various heterogeneous parametric results of studied clusters are found in the literature, which is described as following. The center coordinates of TURNER 6 is given by \cite{fro} as $10^h59^m25.1^s, -59^d32^m12^s$, which are very close to center coordinate given by \cite{dias2} catalogue and \cite{khar} catalogue for TEUTSCH 106 such as $10^h59^m25.0^s,-59^d32^m49^s$ and $10^h59^m25.1^s,-59^d32^m06^s$ respectively. Similarly, the center coordinates of TURNER 6 are extracted from both catalogs follow as, $10^h59^m01.0^s,-59^d29^m58^s$ and $10^h59^m01.0^s,-59^d28^m59^s$ respectively. Moreover, the estimated cluster radius of Turner 6 and Teutsch 106 is given to be 1.65 \cite{bon,dias2} arcmin and 2.60 \cite{ta09,dias2} arcmin respectively.  The center, core-radius and radius of ESO-65SC03 has also been estimated by \cite{khar} as $12^h51^m18^s.0,-69^d43^m48^s.0$, 1.5 arcmin and 8.4 arcmin respectively, whereas \cite{dias2} reported that the center and radius of this systems are to be $12^h51^m37^s.0,-69^d43^m48^s.0$ and 3 arcmin respectively. \cite{jo15} have been declared that ESO-65SC03 is a cluster remnant. According to their analysis, the center of this system is found to be $\alpha = 12^{h}:51^{m}:19.7^{s}, \delta = -69^{o}:43^{'}:21.6^{''}$. All these results of studied clusters are listed in Table~\ref{table}.\\
\begin{table}
\tiny
\caption{The different literature coordinates of the clusters.}
\begin{center}
\begin{tabular}{@{}|c|c|c|c|@{}}
\hline
Literature  & ESO65-SC03 & TURNER 6  & TEUTSCH 106 \\
            & (RA {\&} DEC) & (RA {\&} DEC) & (RA {\&} DEC) \\
\hline%
\cite{khar}   &   12:51:18 & 10:59:01  &  10:59:25.1 \\
              & -69:43:48 & -59:28:59 &   -59:32:06\\ 
\\
\cite{dias2}    &  12:51:37   & 10:59:01 & 10:59:25  \\
                & -69:43:48  & -59:29:58 & -59:32:49  \\
\\
\cite{jo15}    & 12:51:19.7  &   ---      &  ---   \\             
               & -69:43:21.6 &   ---      &  ---   \\
\\
\cite{fro}    &     ---     & 10:59:25.1 &  ---   \\
              &     ---     & -59:32:12  &  ---   \\                 
\hline
\end{tabular}
\end{center}
\label{table}
\end{table}
\section{Stellar sequences and membership of the clusters}
\label{ce03}
The stellar distribution of CMD is an excellent tool to identify the stellar sequences. The stellar evolution process depends on their initial mass and identical to similar luminous stars of cluster. Field star sequences (FSS) is also associated with the MS of OSCs. Moreover, the stellar density of FSS is limited to some magnitude range towards to the fainter end. These field stars can be separate from the members of cluster through the statistical approach \citep{jho} and membership criteria on the kinematic probabilities \citep{zha, yad}. In the statistical approach, the field stars decontamination is carried out according to their colour and magnitude dissimilarities from the members of OSC. The membership probability also depends on the proper motion of these members. \citep{jo15} are reported that the field sequence may still present in CMDs after field star decontamination through the statistical cleaning approach. The presence of field stars is also influenced the stellar distribution and estimation precision of cluster's parameters. In some cases, the stellar density of two identified sequences of an OSC is so much similar that it is difficult to decide actual MS. Since, such type pattern of stellar sequences can also be produced by the field stars, therefore, the well known catalogued clusters is studied by us. We have been discussed some examples of OSC in previous section, in which two stellar sequence is noticed in a single $ (J-H) $ vs $K$ CMD of a cluster. In addition, the  parameters of OSC are estimated through the fitting of theoretical isochrones on these CMDs. These model fit is/may bent at the fainter limit of stellar magnitudes leads a false estimation of distance and reddening values of the cluster due to the overlapping of different stellar sequences. Thus, a new approach is required to distinguish these stellar sequences before estimating the cluster's parameters. Since, the normal CMD does not suitable for this purpose, therefore, CMRD method \citep{jo16} has been applied.\\ 
\begin{figure}[tbp]
\includegraphics[width=25pc]{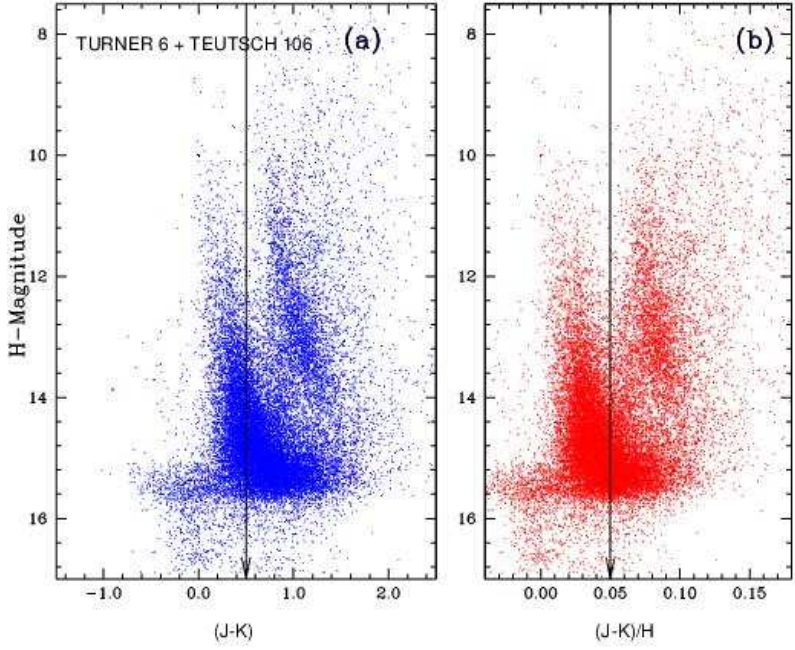}\\
\\
\caption{The upper and lower panels are represented the $ (J-K)$ vs $H$ CMD and $ (J-K) /H$ vs $H$ CMRD for dense cluster regions of $TURNER~6+TEUTSCH~106$. The arrows in the left panel are represented the value of (J-K) (0.5 for TURNER 6 + TEUTSCH 106). Similarly, the arrow in right panel represents the values of CMR i.e. (J-K) /H (0.05 for TURNER 6 + TEUTSCH 106 region). The arrow of the right panel is used to separate the star sequences within the dense cluster region.}
\label{01_cmrd}
\end{figure}
\begin{figure}
\begin{center}
\includegraphics[width=20pc, height=20pc, angle=0]{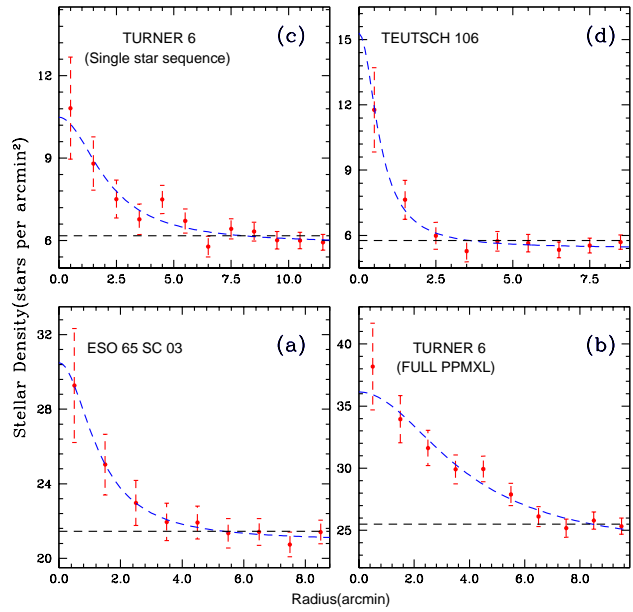}
\caption{The radial density profile (RDP) of clusters through stellar density within concentric rings. The blue dashed line shows the Modified king empirical profile while black dashed line shows the field density. The error of density estimation is calculated by the relation $\sigma_{i}=\frac{\sqrt{N_i}}{A_i}$ in which $N_{i}$ is the total number of stars present an area $A_{i}$}  \label{Fig2}
\end{center}
\end{figure}
In the present case of TEUTSCH 106 and TURNER 6 (T106+T6), the stellar sub-sequence seems to be less dense compared to the MS, therefore, the CMR diagram (CMRD) method is applied for analyzing to it. In the CMRD, x-axis and y-axis represent to $H$ magnitude of stars and corresponding $(J-K)/H$ ratio respectively. In figure\ref{01_cmrd}, the CMD and CMRD are plotted for TURNER 6 in left and right panels, respectively. The CMRD contains two stellar sequences, which are distinguished by a constant colour-magnitude ratio (CMR) value, i.e., 0.05. This value (0.05) is marked by a black arrow in figure \ref{01_cmrd}. It is clearly shown separated sequences of T106+T6 region on the CMD and CMRD diagrams. Thus, CMRD is an excellent tool to estimate the cluster parameters. The advantage of CMRD are given below,\\
(i) The membership uncertainty of stars of OSC is increased with magnitude due to fact that stars of higher magnitude are showing more scattering on the CMD plane. Such scattering is relatively low for CMRD.\\
(ii) The CMRD plane shows more clear separation for stellar sequences compare to CMD.\\
(iii) It is clear from figure 1 that the separation magnitude range is increased in the CMRD compare to CMD. The separation-range of stellar sequences is 7.6-$12.8{\pm}0.4$ H-mag on the plane of CMD, whereas this range increases till to $14.4{\pm}0.4$ H-mag for CMRD.\\
(iv) A solution of linear equation is needed to separate the stellar sequences of CMD, whereas this separation is possible through a constant value of the colour-magnitude ratio of CMRD.
\section{Parameters estimation}
\label{pa04}
The spatial characteristic and dynamical behavior of a cluster is understand through the precise estimation of cluster's parameters such as distance, age, radius, mean proper motion and reddening. These parameters of present studied clusters are estimated through the extracted data from the PPMXL catalog \cite{ros}. This catalog provides the stellar RA-DEC coordinates. The estimation procedure of various parameters of studied clusters are given below,
\subsection{Center, core radius, radius and limit radius}
The radial size of an OSC is defined in the terms of projected radius. The radius can be determined through the radial density profile (RDP) of the cluster. The exact RDP and accurate center of any OSC are not obtained due to the irregular shape and non-uniform distribution of stars at different brightness levels \citep{jho}. The cluster's center holds maximum stellar density in whole cluster region and identified through the stellar distribution of OSCs on $RA-DEC$ plane. Similarly, the RDP of each cluster is constructed for calculating their radius. For this purpose, the cluster region is divided into concentric rings from the cluster's center and the stellar density of each ring is determined by dividing the total number of stars in the ring by its area. The RDP of each cluster is obtained by the fitting of a model \citep{kal} on the RDP of a cluster. The projected radial density $\rho (r) $ of each cluster is estimated by following model:
\begin{equation}
\rho (r)~=~\rho (0) + \frac{f_{0}}{1 + (\frac{r}{r_c})^2}
\end{equation}
where ($r_c$) is the core radius of cluster. This core radius is defined as the radial distance from the cluster's center where the stellar density, reduce half of the central/peak density $f_0$ above the background stellar density $\rho (0) $. The constructed RDP of each cluster is shown in figure 2 (as obtained through PPMXL catalogue). The straight line of each panel is represented the $\rho (b) $, which is intercepted to the RDP of each cluster. The distance of this intercept point from the origin (cluster center) is defined as the cluster's radius. The continuous decrements of stellar density is found in RDP of the cluster from origin to intercept point and no stellar density decrements appears after this point. The stellar density after this point is known as the borderline background stellar density ($\rho_{b}$), which is co-related to the model $\rho (0) $ by a following mathematical expression,
\begin{equation}
\rho_{b}=\rho_{0}+3 \sigma_{bg}
\end{equation}
where $\sigma_{bg}$ is the uncertainty in the estimation of $\rho_{0}$. However, the cluster extent is also determined in the terms of model limiting radius ($r_{limit}$). This radius represented the weakly gravitational bounded region of the cluster \citep{jo15} and estimated by the following relation \citep{buk}:
\begin{equation}
r_{limit}=r_{c} \sqrt{\frac{f_{0}}{3 \sigma_{bg}}-1}.
\end{equation}
Furthermore, this limit radius is used to estimate the cluster's concentric parameter ($c$) by following relation \citep{pet}
\begin{equation}
c=log \left(\frac{r_{limit}}{r_c} \right)
\end{equation} 
Above these prescribed parameters of each cluster are briefly described as following and also summarized in table \ref{tab1}. In this table, the peak density $f_{0}$ is the difference between the value of the y-intercept of the model and the background stellar density $\rho_{0}$.
\subsubsection{ESO-65SC03}
The new center-coordinates are ($12^{h}51^{m}11^{s}.5,-69^{d}42^{m}34^{s}.2$) with the uncertainty of 5 arcsec. The RDP of this cluster is constructed from the data-points of stellar densities of concentric ring of 1.0 arcmin width. The core radius, radius and limit radius of cluster are found to be $1.3{\pm}0.2$ arcmin, $5.0{\pm}0.2$ arcmin and $5.48$ arcmin respectively. Its RDP plot is constructed through the PPMXL catalogue, which provides different values compare to work of 2MASS catalogue \cite{jo15}. 
\begin{table}
\tiny
\caption{The various spatial parameters such as, radius, limit radius, etc. have been summarized in this table.}
\begin{center}
\begin{tabular}{@{}|c|c|c|c|@{}}
\hline
Parameter$\downarrow$/Cluster$\rightarrow$ &ESO65-SC03 & Teutsch 106 & Turner 6\\\hline
RA(center) & $12^{h}51^{m}11^{s}.5$ & $10^{h}59^{m}23^{s}.3$ & $10^{h}58^{m}47^{s}.1$ \\
DEC(center) & $-69^{d}42^{m}34^{s}.2$ &$-59^{d}32^{m}36^{s}.9$ & $-59^{d}34^{m}22^{s}.1$\\
$r_c$ (arcmin) & 1.30 & $0.69{\pm}0.23$ & $2.11{\pm}0.58$  \\
$\rho_{0}$ & 20.91 & $5.41{\pm}0.12$ & $5.87{\pm}0.10$  \\
 $\sigma_{bg}$ & 0.17 & 0.12 & 0.10 \\
$\rho_b$ & 21.42 & 5.77 & 6.17 \\
Peak density & $9.58{\pm}1.52$ & $9.89{\pm}3.53$ & $4.61{\pm}1.25$ \\
Radius (arcmin) & 5.0 & 3.5 & 8.5 \\
$r_{limit}$ (arcmin) & 5.48 & 3.55 & 9.09 \\ 
$c$ & 0.62 & 0.71 & 0.63 \\
\hline
\end{tabular}
\end{center}
\label{tab1}
\end{table}
\subsubsection{TURNER 6 and TEUTSCH 106}
The RDP of this cluster region is constructed through the extracted data from the $PPMXL$ catalog. The cluster's center is found to be ($10^{h}59^{m}19^{s}.0$, $-59^{d}32^{m}20^{s}.9$) in $RA-DEC$ plane through the maximum count method \citep{joshi+2015} which is closed to the coordinates given by \cite{fro} and \cite{dias2} for Turner 6 and Teutsch 106 respectively. In addition, the core radius and radius of this cluster is estimated to be $4.18{\pm}0.77$ arcmin and 8.5 arcmin respectively. Thus, computed radius is found to be higher than the given radius in literature. In addition, the different stellar sequences (as founded by CMRD approach) are also used to this purpose. The stellar sequence (having a CMR value less than 0.5) referred as the TEUTSCH 106 due to the most recent naming of clusters and its results. The center, core-radius, radius and limit radius for TEUTSCH 106 are found to be ($10^{h}59^{m}23^{s}.3$, $-59^{d}32^{m}36^{s}.9$), $0.69{\pm}0.23$ arcmin, $3.5$ arcmin and $3.55$ arcmin respectively. The second stellar sequence referred as the TURNER 6 and its best RDP is obtained by those stars which are spread out in whole cluster region. The center, core-radius, radius and limit radius of TURNER 6 are found to be ($10^ {h} 58^ {m} 47^ {s}. 1, -59^ {d} 34^ {m} 22^ {s}.1$), $2.11{\pm}0.58$ arcmin, $8.5$ arcmin and $9.09$ arcmin respectively.\\
We are also constructed the RDP plot of TURNER 6 (Full region) through the PPMXL catalog, which provides an entirely different set of results. The resultant RDPs are depicted in the figure \ref{Fig2}.
  
\subsection{Probable members and mean proper motion}
Mean proper motion of clusters is an angular displacement rate and it measures in the unit of mili-arcsec (mas) per year. The symbols $\mu_{x}$ and $\mu_{y}$ are represented the stellar proper motion in $RA$ and $DEC$ directions, respectively. The values of stellar proper-motion of the clusters' members are extracted from PPMXL catalog. This catalog is provided a list of about 900 million stars with an accuracy of 80-300 mas in stellar coordinate and 4-10 mas per year in the absolute proper motion and also has the completeness of stars at fainter limit.\\
\begin{table}
\tiny
\caption{The cluster's mean proper motion values in $RA$ and $DEC$ direction using $PPMXL$ and $UCAC4$ have been listed here.}
\begin{center}
\begin{tabular}{@{}|c|c|c|c|@{}}
\hline
Cluster$\downarrow$ &Catalogue$\rightarrow$ & PPMXL & UCAC4  \\
                      &Values$\downarrow$ &&\\
\hline%
{\bf ESO65-SC03} & $\mu_{x}$  & $-8.59{\pm}1.29$ & $-0.72{\pm}0.69$   \\
            & $\mu_{y}$  & $ 5.29{\pm}1.03$ &  $ 0.37{\pm}0.71$  \\\hline
 {\bf TEUTSCH 106} & $\mu_{x}$  & $-0.78{\pm}2.71$ & $-1.13{\pm}0.94$  \\
            & $\mu_{y}$  & $-3.11{\pm}2.74$ & $-0.93{\pm}0.83$  \\\hline
{\bf TURNER 6} & $\mu_{x}$  & $3.53{\pm}1.91 $ & $0.04{\pm}1.26$   \\
            & $\mu_{y}$  & $1.56{\pm}1.89 $ & $0.48{\pm}1.41$  \\\hline 
\end{tabular}
\end{center}
\label{tab2}
\end{table}
\begin{figure}
\begin{center}
\includegraphics[width=18.5pc, height=18.5pc, angle=0]{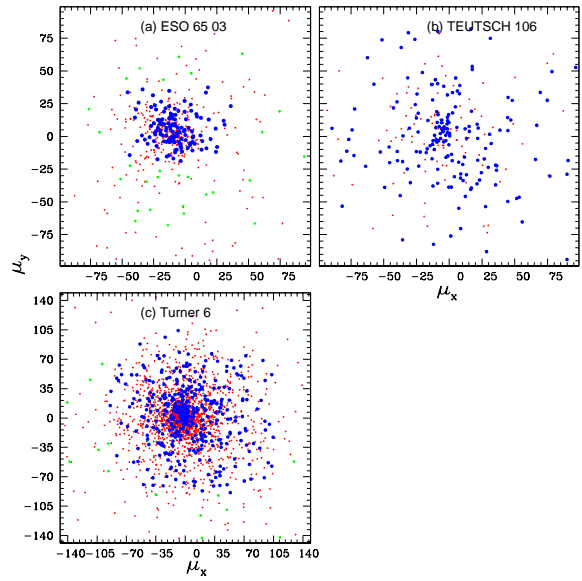}
\caption{The distribution of stars in the ${\mu_{x}-\mu_{y}}$ plane, which have been used to estimate the mean proper motion of the cluster. The red, green and blue dots are depicted the all detected stars, the remaining members through the CMRD separation and final members through the proper motions.}\label{Fig3}
\end{center}
\end{figure}
We are obtained the probable members of the cluster through the combined effort of the CMRD and statistical cleaning approach. In the later approach, the area of field region is taken to be equal to the area of the cluster. Furthermore, we are constructing a grid cell of each field star in the colour-magnitude space. That star of cluster region is rejected from its membership, which is lying in grid-cell of field star and having minimum color-magnitude distance from the star field. Since, the grid size of field stars depends on the stellar density of the cluster, therefore smaller cells are accepted in dense clusters while larger ones for the sparse ones \citep{corr}. These probable members are used to determine the mean proper motion of the cluster by using the iteration method \citep{jho,joshi+2015}. In this method, the mean and standard deviation ($\sigma$) values of proper motion of the cluster's members are determined in both right accession(RA) and declination(DEC) directions and rejected membership of those stars which are laying outside the 3$\sigma$ value of the mean proper motion of any direction. The whole procedure repeated until all stars are not lying within the 3$\sigma$ value of the said mean in both directions. The left members are used to determine the mean proper motion of individual cluster (as depicted in RA-DEC plane by blue dots in figure 3).\\
The proper motions of stars of stellar sequences of OSCs are extracted from the PPMXL catalogue. After applying the CMRD approach, the number of extracted data-points are found to be 450, 285 and 1252 stars for ESO 65, 03, TEUTSCH 106 and TURNER 6 respectively. After applying the statistical cleaning approach, we are found 170, 178 and 422 members for the ESO 65 03, TEUTSCH 106 and TURNER 6 respectively. The iteration method of $\sigma$-clipping algorithm are provided the mean proper motion of 139, 171 and 395 members for ESO 65 03, TEUTSCH 106 and TURNER 6 respectively. The resultant values of mean proper motion are listed in table \ref{tab2}. These results are shown close agreement with the \cite{khar}. The corresponding mean proper values of clusters through UCAC4 catalog are also summarized in table \ref{tab2}.\\
In addition, the mean proper motion for TURNER 6 (full Sky region of cluster) is found to be $\bar{\mu}_{x}=-0.72{\pm}0.69$ and $\bar{\mu}_{y} = 0.37{\pm}0.71$ in RA and DEC directions respectively. These values are determined through the proper motion values of the 1959 members of TURNER 6, which are left after $\sigma$-clipping algorithm on the all 2163 identified members in described region.\\
\subsection{CMD: Distance and Age}
The stellar broadening is found on CMDs due to the presence of probable binaries and field stars \citep{sha}. However, the stars of MS of CMDs are excellent tools to estimate the physical parameters (distance, age, reddening etc.) of clusters. These parameters can be obtained for the clusters by best fitted theoretical isochrones. These isochrones are based on model dependent mass, radius and distance of each star. Since, the near-Infrared surveys of clusters are less affected by high reddening from the Galactic plane, therefore, they are selected to estimate the clusters' parameters. The field star decontamination are already carried out in the previous section. The identified probable members are also utilized to determine the distance and age of the clusters. These parameters are estimated by visual inspection of best fitted isochrones on cluster CMDs, which is obtained by checking the several Padova group isochrones of stellar evolution model \citep{mar,gir}. The distance and age are estimated by keeping the fixed value of colour excess $(J-H)$ and $(H-K)$ on $H$ versus $(J-H)$ and $H$ versus $(H-K)$ CMDs in such a manner that these colour-excesses values should be obeyed \cite{fio}'s relations for normal interstellar medium. The distance of a cluster is estimated by following relation:
\begin{equation}
Distance(kpc)=10^{1+\frac{H-A_{H}}{5}}
\end{equation}
where $H$ is the total cluster H-band magnitude and $A_{H}$ is extinction in $H$ band which has been found by the relation $A_{H}=0.176 A_{V}$ \cite{sch} and $A_{V}=R_{V}E(B-V)=3.1E(B-V)$ \citep{sch77}. The value of reddening i.e., $E(B-V)$,  is estimated by the relation $E(J-H)/E(B-V)=0.31{\pm}0.13$ \citep{fio}. The results of age and distance of each cluster are described as below,\\
\subsubsection{ESO65-SC03}
In the case of present studied cluster, the grid size is taken to be $({\pm}0.3,{\pm}0.06)$ for statistical approach. The log-age, the total cluster $H$-band magnitude, distance modulus and colour-excess (J-H) are found to be $8.75{\pm}0.05$, $12.5{\pm}0.1$, $12.4{\pm}0.1$ and $0.15$ respectively, lead to age, distance and reddening of studied cluster as $0.56{\pm}0.01~Gyr$, $3.04{\pm}0.30~kpc$ and $0.38$ respectively. The estimated age is very close to $3.047~kpc$ as estimated by \cite{khar}. The best fitted isochrone of solar mettallicity is depicted in the figure 4(a).\\  
\begin{figure}
\begin{center}
\includegraphics[width=12.0pc,angle=-90]{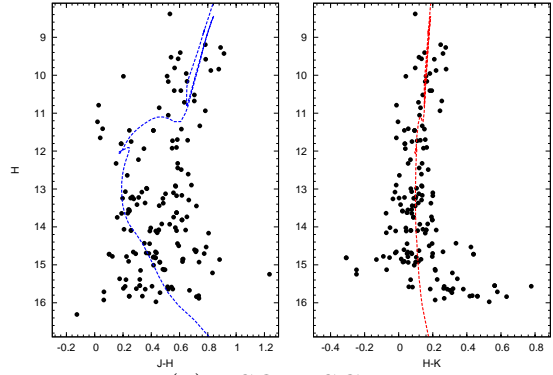}\\
{(a) ESO65-SC03}\\
\includegraphics[width=12.0pc,angle=-90]{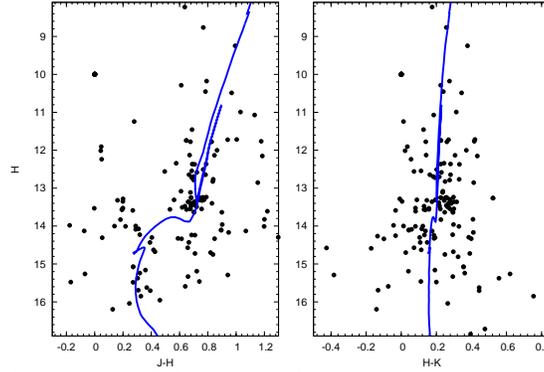}\\
{(b) TEUTSCH 106 (Statistical cleaned)}\\
   \includegraphics[width=12.0pc,angle=-90]{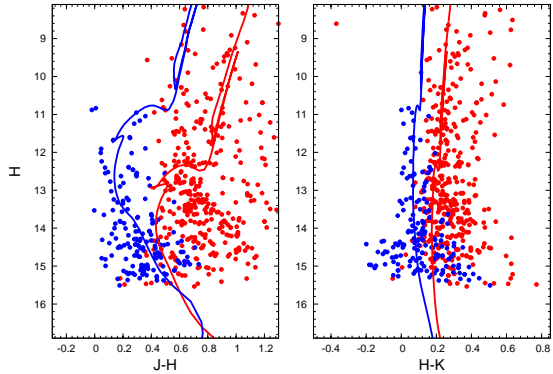}\\
{(c) TEUTSCH 106 and TURNER 6}
\caption{The $(J-H)~ vs~ H$ and $(H-K)~ vs ~ H$ CMDs for the studied clusters. The black dots in each panel (excluding panel (c)) are representing those stars which are remaining after field star subtraction from cluster region. The blue and red dots in panel (c) have represented the stars of TEUTSCH 106 and TURNER 6 respectively.}\label{Fig5}
\end{center}
\end{figure}
\begin{table}
\tiny
\caption{The various parameters obtained through CMDs are summarized here.}
\begin{center}
\begin{tabular}{@{}|c|c|c|c|c|@{}}
\hline
Cluster/Parameter &Distance & Age & Distance & Reddening\\
 &modulus (mag) & (Gyr) &  (kpc) & (mag)\\\hline
ESO65SC03 & $12.42{\pm}0.20$ & $0.56{\pm}0.05$ & $3.04{\pm}0.30$ & $0.38$\\
TURNER 6 & $12.47{\pm}0.1$ & $0.89{\pm}0.10$ & $3.12{\pm}0.03$ & 1.03 \\
TEUTSCH 106 & $11.19{\pm}0.1$ & $0.79{\pm}0.05$ & $1.73{\pm}0.02$ & 0.32 \\
\hline
\end{tabular}
\end{center}
\label{tab3}
\end{table}
\subsubsection{TURNER 6 and TEUTSCH 106}
A typical distance of TEUTSCH 106 is determined through the same cluster region as selected by \cite{dias2}. The field star decontamination of this region is performed by adopting a grid size ($0.2$mag, $0.05$mag) on ($H$, $J-H$) CMD. The distance and reddening of TEUTSCH 106 are determined to be $6.6$ kpc and $0.81$ mag, respectively, which show good agreement with distance $6.7$ kpc and reddening $0.97$ mag as estimated by \cite{khar}. It is clear from figure 4(b) that the probable members of this region are showing broad scatter pattern of stars. These values are obtained by the combined stellar sequences of TURNER 6 and TEUTSCH 106 (hereinafter T106+T6).\\
Other hand, statistical cleaned approach (grid size, $(H,(J-H))={\pm}0.05,{\pm}0.05$) are also used to separate the both identified stellar sequences through CMRD approach. The age, distance, and reddening of TEUTSCH 106 are found to be $0.79{\pm}0.05$ Gyr, $1.73$ kpc and $0.32$ mag respectively. These results are far away from the results of \cite{khar} and \cite{dias2}. Our selected TEUTSCH 106 region and T106+T6 region are overlap to each other but have different size. The statistical cleaned CMD of T106+T6 is not showing different stellar sequences due to the higher rejection of stars by larger grid size. The larger grid size is against of the hypothesis of small grid size of dense region. Moreover, larger grid size does not reduced the stellar scattering in CMD. The broaden and scattered distribution of stars can become the cause of wrong estimation of visual fit of isochrones on CMD. Since, the parametric results of TEUTSCH 106 is not matched with the said T106+T6 region, therefore the analysis of stellar sequences of cluster regions may open a new window of cluster studies.\\
The distance and reddening of TURNER 6 are estimated to be $3.12$ kpc and $1.03$ mag respectively. \cite{khar} are also estimated the distance and reddening of this cluster as $3.6$ kpc and $1.14$ mag, respectively. Our present results are showing good agreement with the results of \cite{khar}. Similarly, the distance of studied cluster is estimated to be $3.25$ kpc by \cite{dias2}, which is close to our results. The red and blue lines are representing the best fit isochrone of solar and 0.008 metallicity, respectively and depicted in figure 4(c).\\
The summary of distance, age and reddening is summarized in the table \ref{tab3}.
\begin{figure}
\begin{center}
\includegraphics[width=20pc, height=30pc, angle=270]{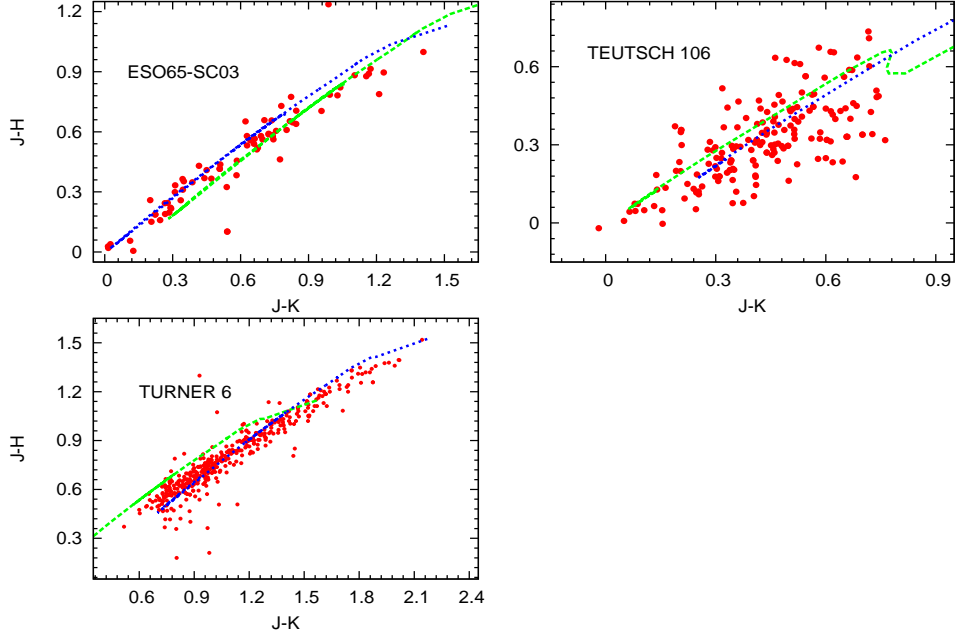}
\caption{The $(J-K)/(J-H)$ two-colour diagram for present studied clusters.}\label{Fig6}
\end{center}
\end{figure}
\begin{table}
\tiny
\caption{The slopes of the $(H-{\lambda})$ versus $(H-K)$ diagrams in the direction of cluster.}
\begin{center}
\begin{tabular}{@{}|c|c|c|c|@{}}
\hline
Ratios $/$ Cluster & ESO65-SC03  & TURNER 6  & TEUTSCH 106  \\
\hline%
$\frac{(H-B_{ph})}{(H-K)}$ & $-10.57{\pm}1.23$ & $-6.66{\pm}0.88$ & $4.97{\pm}2.58$ \\
$\frac{(H-R_{ph})}{(H-K)}$ & $-7.17{\pm}1.17$  & $-7.46{\pm}0.75$ & $0.50{\pm}3.94$ \\ 
$\frac{(H-I_{ph})}{(H-K)}$ & $-4.74{\pm}0.66$  & $-5.09{\pm}0.94$ & $-1.67{\pm}2.04$\\
$\frac{(J-H)}{(H-K)}$      & $1.54{\pm}0.19$   & $1.80{\pm}0.09$  & $0.10{\pm}0.32$ \\
$\frac{(H-W_1)}{(H-K)}$    & $0.89{\pm}0.13$   & $1.46{\pm}0.10$  & $0.86{\pm}0.28$ \\
$\frac{(H-W_2)}{(H-K)}$    & $0.76{\pm}0.18$   & $1.53{\pm}0.11$  & $0.66{\pm}0.42$ \\
\hline
\end{tabular}
\end{center}
\label{tab4}
\end{table}
\subsection{Two colour diagrams (TCD) and ratios}
Two-colour diagrams(TCD) are useful to determine the reddening and investigate the nature of reddening law in the direction of cluster region. The normal reddening law $R_V~=~\frac{A_V}{E(B-V)}$ is valid for lines of sights which do not pass through the dense clouds \citep{sne}.
\subsubsection{Reddening through $(J-K)/(J-H)$ diagram}
To determine interstellar extinction in the near-IR, the $(J-K)/(J-H)$ diagram is used and depicted in the figure \ref{Fig6}. We applied the normal reddening law for the infrared colours and shifted the stars along the vector $\frac{E(J-H)}{E(J-K)}$ using different theoretical isochrones \citep{mar}. The relation $E(J-K)/E(B-V)=0.72{\pm}0.05$ \citep{mor} is used to estimate the reddening i.e., $E(B-V)$, for individual cluster. The colour-shifts ($\Delta_{(J-H)}$, $\Delta_{(J-K)}$) of ESO65-SC03, TEUTSCH 106 and TURNER 6 are found to be (0.26, 0.15), (0.19, 0.12) and (0.61, 0.38) respectively. These shifted values are used to determine the reddening values of ESO65-SC03, TEUTSCH 106 and TURNER 6. The reddening values are found as 0.36, 0.26 and 0.84 mag respectively. Moreover, these reddening values are showing good agreement with the values of table \ref{tab3}.\\
\subsubsection{Colour-ratios}
The colour ratio values of two colour-diagrams are important for investigating the nature of extinction law, dependency with each other and their variation from normal values \citep{jo16}. The linear relationship of various colours [$(H-{\lambda})$, where $\lambda=B_{ph},R_{ph},I_{ph},J,W1$ and $W2$ bands] of each cluster are estimated with the colour $(H-K)$. The best linear fit values are listed in table 5. The colour-ratio results of $TEUTSCH~106$ are found to be abnormal compare to other clusters. These abnormal values are interesting puzzle to understand the cause of such type variation. These variation may be a result of present of molecular cloud/dust, which is also supplemented by mid-IR colours. Furthermore, the values of these ratios are found normal for other studied clusters. In addition, $(H-W_1)/(H-K)$ values for ESO65-SC03 and TEUTSCH 106 are found more than $(H-W_2)/(H-K)$ whereas reverse fact is found for TURNER 6.
\section{Two-colour Magnitude Ratio Diagrams}
\label{tc05}
The colour-magnitude ratio values of CMRD are more effective compare than the colour values of CMD. The stellar scattering of CMD plane is greater than CMRD plane. In this background, a new two-colour magnitude ratio diagram (TCMRD) approach is proposed to investigate the nature of extinction law and to reduce the stellar scattering of TCD. In this approach, the colour values of an individual star are normalized with respect to its magnitude. The normalized colour-ratio values are different from the normal colour ratio values, which is clearly verified through the table 5 and 6. The resultant TCMRD of clusters are shown in the figure 6.\\  
\begin{table}
\tiny
\caption{The slopes of the $(H-{\lambda})/H$ versus $(H-K)/H$ diagrams in the direction of the cluster.}
\begin{center}
\begin{tabular}{@{}|c|c|c|c|@{}}
\hline
Ratios $/$ Cluster & ESO65-SC03 & TURNER 6  & TEUTSCH 106 \\
\hline%
$\frac{(H-B_{ph})/H}{(H-K)/H}$ & $-16.05{\pm}1.09$ & $-10.91{\pm}0.76$& $4.93{\pm}2.67$\\
$\frac{(H-R_{ph})/H}{(H-K)/H}$ & $-11.67{\pm}1.04$ & $-8.77{\pm}0.58$ & $-0.71{\pm}4.13$\\ 
$\frac{(H-I_{ph})/H}{(H-K)/H}$ & $-6.61{\pm}0.52$  & $-6.43{\pm}0.68$ & $-1.83{\pm}2.26$\\
$\frac{(J-H)/H}{(H-K)/H}$      & $2.09{\pm}0.15$   & $2.06{\pm}0.08$  & $0.07{\pm}0.34$ \\
$\frac{(H-W_1)/H}{(H-K)/H}$    & $1.08{\pm}0.09$   & $1.54{\pm}0.07$  & $0.85{\pm}0.29$ \\
$\frac{(H-W_2)/H}{(H-K)/H}$    & $0.87{\pm}0.12$   & $1.49{\pm}0.08$  & $0.73{\pm}0.43$ \\
\hline
\end{tabular}
\end{center}
\label{tab5}
\end{table}
\begin{figure}[tbp]
\centering
\includegraphics[width=13.0pc,angle=270]{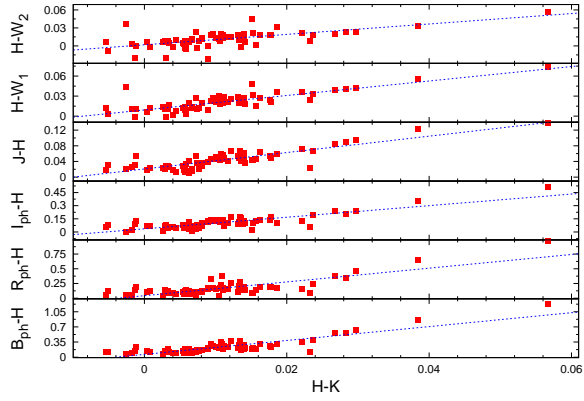}\\
{(a) ESO65-SC03}\\
\includegraphics[width=13.0pc,angle=270]{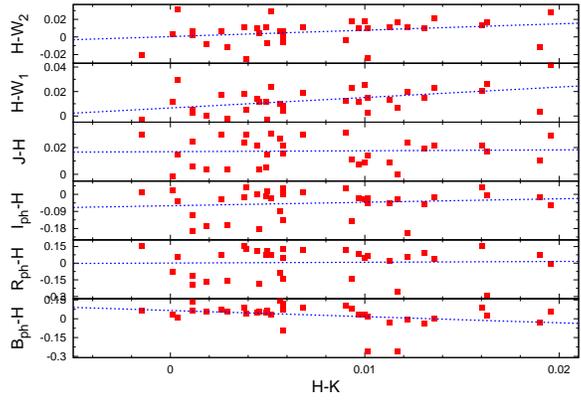}\\
{(b) TEUTSCH 106}\\
\includegraphics[width=13.0pc,angle=270]{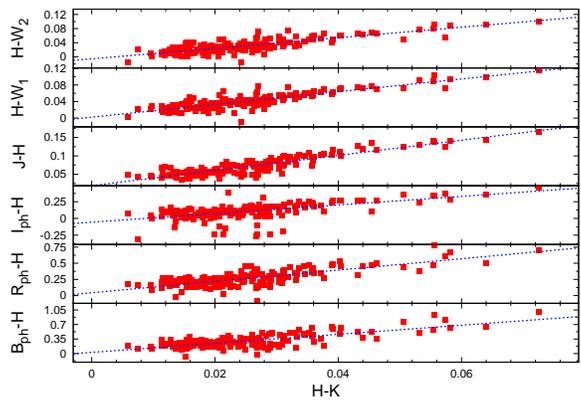}\\
{(c) TURNER 6}
 \hfill
\caption{\label{Fig7} The two-colour-magnitude ratio diagrams of cluster have shown here.}
\end{figure}
The difference of normalized colour ratio and normal colour ratio seems to be high in the shorter wavelength ($\lambda$) whereas this difference gradually decreases toward to infrared wavelength. Both ratio values seem to be identical for $W_2$ band. In the case of TEUTSCH 106, said ratio may be arising due to the stellar sub-sequence. To verify the reality of prescribed ratios of the TEUTSCH 106, the whole cluster region is adopted to estimate the colour ratios. Since, the resultant colour-ratios are abnormal, therefore, the sub-sequence is not responsible for the abnormal values of colour-ratios. Moreover, colour-ratio values of TEUTSCH 106 are abnormal for any one of $I_{ph}$, $B_{ph}$, $R_{ph}$ and J band, whereas the similar values of TCR and TCMR are obtained for $K$, $W_1$ and   $W_2$ bands. These deviated values of TEUTSCH 106 region due to the large quantity of stellar dust and gas clouds within it. It seems through table 5 and 6 that the uncertainties of CMR values are decreased for ESO 65 03 and TURNER 6 whereas in the case of TEUTSCH 106, uncertainties of CMR values are increasing according to normal color values. On this background, it is concluded that the stellar scattering is effectively reduced in TCMRD planes for the dust free cluster region, whereas it is increased for dust-affected cluster regions. Thus, the stellar scattering of TCMRD is directly depended to the amount of stellar dust.
\section{Young Stellar Object Fraction}
\label{yo06}
The number of Young stellar objects (YSOs) within OSCs is dependent on its age, the YSO fraction of any cluster reduces with its evolution. We are determined the free reddening parameter of each star by following expression of \cite{buc}
\begin{equation}
Q=(J-H)-\frac{E(J-H)}{E(H-K)} \times (H-K)
\end{equation} 
The value of E(J − H)/E(H − K) = 1.55 \cite{mat} is used for all determinations of Q. YSOs are stellar objects having Q value less than -0.5 mag \citep{buc}. The YSOs' fraction of a cluster is a ratio of the number of YSOs ($N_ {YSO} $) and the total number of cluster members ($N_ {cl} $). To reduce the photometric scattering in small number of YSOs within cluster, the YSOs fraction is estimated by following relation \citep{buc}:
\begin{equation}
Y_{frac}=\frac{N_{YSO}-\sqrt{N_{YSO}}}{N_{cl}},
\end{equation} 
There are 7, 3 and 94 YSOs are identified for $ESO65 SC 03$, $TEUTSCH~106$ and $TURNER 6$ regions, respectively. Similarly, the $Y_{frac}$ of these clusters are found to be 0.026, 0.000 and 0.003 respectively. After comparison of our YSO fraction values with the ``APPENDIX A: FSR CLUSTER PROPERTY TABLE" of the manuscript of \cite{buc}, we are concluded that the low fractional values are found for $TEUTSCH~106$ and $TURNER 6$ regions, whereas the average fraction value is obtained for the ESO65 SC 03.
\section{Discussion and Conclusion}
\label{co07}
The OSCs are good tracer of galactic evolution process and star formation history. These studied are possible due to the precise measurement of their parameters. Since, the field stars (FS) of OSCs are reduced the precision in estimation of parameters, therefore their decontamination (separation of cluster members) are needed. The said decontamination is carried out through the statistical cleaning approach \citep{jo15}, which is based on the colours and magnitudes of the stars. Since, this approach is highly dependent on the adopted grid size around field stars; therefore the decontamination of FS required some more attention. In addition, the identification of MS of the statistical cleaned ($J-H$)/$H$ CMD (refer figure 4-b) is difficult due to non-fixed separation boundary of fainter stars within the CMD plane. This difficulty can overcome through the CMRD approach. The separated stellar sequences are used to identify the MS of the cluster (ex. TEUTSCH 106). Such analysis of stellar sequences of Sky region will open a window of identification of new clusters within embedded region of the Galactic disk and highly influenced region by the interstellar gases. We are summarized our results of the spatial morphological parameters of each cluster in table \ref{tab1}. We are found different results of spatial parameters of studied clusters through the extracted data sets from PPMXL catalogue and 2MASS catalogue. Such parametric variation arises due to the fact that PPMXL catalogue is contained a wide number of fainter stars compare than 2MASS catalogue. The multi-band catalogue is seems to be more effective than the catalogue through the one or two bands. The PPMXL catalogue also contains some false fainter stars, therefore, it is needed to verify the stellar nature before using it. On this background, we are mostly used 2MASS catalogue for parametric analysis of the present studied clusters.\\ 
The distance, age and reddening of each cluster have been estimated by best fitted theoretical isochrones (based on model star radius, temperature and chemical composition) on $J-H$ vs $H$ and $H-K$ vs $H$ CMDs. These results are summarized in table \ref{tab3}. The resultant reddening values are obtained by $(J-K)/(J-H)$ TCD and these values are close to reddening values of table \ref{tab3}. The cluster's members are changed their angular position leads to as their proper motion. Since, the each member of the cluster has different proper motion values, therefore, mean proper motion of the cluster would useful to understand the cumulative properties of members and these values of each cluster are summarized in table 3. The nature of extinction law is varied within OSCs due to the  presence of interstellar gaseous/clouds/dust. The clour-ratio of clusters are determined through the linear fits and listed in table \ref{tab4}, whereas normalized colour-ratios have been summarized in table \ref{tab5}.\\
The separated stellar-sequences of clusters provide the crucial information about their evolution process. The constant value of normalized-colour ratios are used to separate the stellar sequences on CMRD plane. The straight dividing line and stellar separation of the fainter stars are the main features of the CMRD, which is not possible on normal CMD plane. The variation between the values of normalized colour-ratio and normal colour ratio is higher for the shorter wavelength, whereas it is decreased toward to the higher wavelength. The TCD and TCMRD are indicated that the ratio values of TEUTSCH 106 are far away from the values of other clusters. It may possible due to the presence of inter-stellar dust/gases within the field view of the TEUTSCH 106. The stellar scattering of TCMRD plane depends on the amount of associated interstellar-gaseous. Thus, the study of stellar sequences will become a milestone to identify the new clusters. The stellar enhancement does not only a single property of the cluster, but an associate stellar sequence is more important to understand the spatial and dynamical morphology of the cluster. We are found greater size of TEUTSCH 106 compare than \cite{dias2}.  
\section*{Acknowledgments}
GCJ is thankful to AP-Cyber Zone (Nanakmatta) for providing computer facilities. Author is also acknowledge the SIMBAD and VIZIER services. This publication made use of data products from the Two Micron All Sky Survey, which is a joint project of the University of Massachusetts and the Infrared Processing and Analysis Center/California Institute of Technology, funded by the National Aeronautics and Space Administration and the National Science Foundation. This publication is also makes use of data products from the Wide-field Infrared Survey Explorer, which is a joint project of the University of California, Los Angeles, and the Jet Propulsion Laboratory/California Institute of Technology, funded by the National Aeronautics and Space Administration.


\begin{thebibliography}{}

\bibitem[\protect\citeauthoryear{Bate et al.}{2003}]{bat}
Bate, M. R., Bonnell, I. A. \& Bromm, V., 2003, MNRAS, 339, 577

\bibitem[\protect\citeauthoryear{Bonatto \& Bica}{2010}]{bon}
Bonatto, C., Bica, E., 2010 A{\&}A 521, A74

\bibitem[\protect\citeauthoryear{Bonatto et al.}{2006}]{bo06}
Bonatto, C., Bica, E., Ortolani, S. and Barbuy, B., 2006, A{\&}A, 453, 121

\bibitem[\protect\citeauthoryear{Buckner \& Froebrich}{2013}]{buc}
Buckner, A. S. M. \& Froebrich D., 2013,  MNRAS, 436, 1465

\bibitem[\protect\citeauthoryear{Bukowiecki et al.}{2011}]{buk}
Bukowiecki, ${L}$, Maciejewski, G., Konorski, P. and Strobel, A., 2011, AcA, 62

\bibitem[\protect\citeauthoryear{Corradi et al.}{2009}]{corr}
Corradi, W. J. B., Maia, F. F. S. and Santos-Jr., J. F. C., 2009, Star Clusters - Galactic Building Blocks throughout Time and Space, Proceedings IAU Symposium No. 266

\bibitem[\protect\citeauthoryear{Dias et al.}{2002}]{dias1}
Dias, W. S., Alessi, B. S., Moitinho, A. and Lepine, J. R. D., 2002, A{\&}A, 389, 871

\bibitem[\protect\citeauthoryear{Dias et al.}{2012}]{di12}
Dias, W. S., Monteiro, H., Caetano, T. C. and Oliveria, A. F., 2012, A{\&}A, 539, A125

\bibitem[\protect\citeauthoryear{Dias et al.}{2014}]{dias2}
Dias, W. S., Monteiro, H., Caetano, T. C., Lepine, J. R. D., Assafin, M. and Oliveria, A. F., 2014, A{\&}A, 564, A79

\bibitem[\protect\citeauthoryear{Fiorucci \& Munari}{2003}]{fio}
Fiorucci, M. \& Munari, U. 2003, A{\&}A, 401, 781

\bibitem[\protect\citeauthoryear{Friel}{1995}]{fri}
Friel, E. D., 1995, ARA{\&}A, 33, 381

\bibitem[\protect\citeauthoryear{Froebrich}{2010}]{fro}
Froebrich, 2010, MNRAS, 409, 128

\bibitem[\protect\citeauthoryear{Girardi et al.}{2010}]{gir}
Girardi, L. et al., 2010, ApJ, 724, 1030

\bibitem[\protect\citeauthoryear{Harris \& Pudritz}{1994}]{hap}
Harris, W. E. \& Pudritz, R. E., 1994, AJ, 429, 177

\bibitem[\protect\citeauthoryear{Hasan}{2005}]{ha05}
Hasan, P., 2005, Bull. Astr. Soc. India, 33, 151

\bibitem[\protect\citeauthoryear{Hasan et al.}{2008}]{has}
Hasan, P., Kilambi, G. C. \& Hasan, S. N., 2008, Astrophysics and space science, 313, 363

\bibitem[\protect\citeauthoryear{Joshi et al.}{2014}]{jho}
Joshi, Y. C., Balona, L. A., Joshi, S. \& Kumar, B., 2014, MNRAS, 437, 804

\bibitem[\protect\citeauthoryear{Joshi et al.}{2015a}]{jo15}
Joshi, G. C., Joshi, Y. C., Joshi, S., Chowdhury, S. \& Tyagi, R. K., 2015, PASA, 32, 22

\bibitem[\protect\citeauthoryear{Joshi et al.}{2015b}]{joshi+2015}
Joshi, G. C., Joshi, Y. C., Joshi, S. \& Tyagi, R. K., 2015, New Astronomy, 40, 68

\bibitem[\protect\citeauthoryear{Joshi \& Tyagi}{2015}]{jr15}
Joshi, G. C. \& Tyagi, R. K., 2015, 6$^{th}$ International Conference On “Recent Trends in Applied Physical, Chemical Sciences, Mathematical/Statistical and Environmental Dynamics” (PCME-2015), pp.37-42

\bibitem[\protect\citeauthoryear{Joshi \& Tyagi}{2016}]{jo16}
Joshi, G. C. \& Tyagi, R. K., 2016, MNRAS, 455, 785

\bibitem[\protect\citeauthoryear{Kaluzny \& Udalski}{1992}]{kal}
Kaluzny, J. \& Udalski, A., 1992, AcA, 42, 49

\bibitem[\protect\citeauthoryear{Kharchenko et. al.}{2013}]{khar}
Kharchenko, N. V., Piskunov, A. E., Schilbach, E., Ro${\ddot{e}}$ser, S. and Scholz, R.-D., 2013, A{\&}A, 558, A53

\bibitem[\protect\citeauthoryear{Marigo et al.}{2008}]{mar}
Marigo, P. et al., 2008, A{\&}A, 482, 883

\bibitem[\protect\citeauthoryear{Mathis}{1990}]{mat}
Mathis, J. S., 1990, ARA{\&}A, 28, 37 

\bibitem[\protect\citeauthoryear{Morgan \& Nandy}{1982}]{mor}
Morgan, D. H. \& Nandy, K., 1982, MNRAS, 199, 979

\bibitem[\protect\citeauthoryear{Peterson \& King}{1975}]{pet}
Peterson, C. J. \& King, I. R., 1975, AJ, 80, 427

\bibitem[\protect\citeauthoryear{Ro$\ddot{e}$ser et al.}{2010}]{ros}
Ro${\ddot{e}}$ser, S., Demleitner, M. \& Schilbach, E., 2010, AJ, 139, 2440

\bibitem[\protect\citeauthoryear{Santos Jr. et al.}{2005}]{sa05}
Santos Jr., J. F. C., Bonatto, C. and Bica, E., 2005, A{\&}A, 442, 201

\bibitem[\protect\citeauthoryear{Schlegel et al.}{1998}]{sch}
Schlegel, D., Finkbeiner, D. \& Davis, M., ApJ, 1998, 500, 525

\bibitem[\protect\citeauthoryear{Sharma et al.}{2006}]{sha}
Saurabh, S., Pandey, A. K., Ogura, K., Mito, H., Tarusawa, K. \& Sagar, R., 2006, AJ, 132, 1669

\bibitem[\protect\citeauthoryear{Schild}{1977}]{sch77}
Schild, R., 1977, AJ, 82, 337

\bibitem[\protect\citeauthoryear{Skrutskie et al.}{2006}]{skr}
Skrutskie, M. F., et al., 2006, AJ, 131, 1163

\bibitem[\protect\citeauthoryear{Sneden et al.}{1978}]{sne}
Sneden, C., Gehrz, R. D., Hackwell, J. A., York, D. G., Snow, T. P., 1978, APJ, 223, 168

\bibitem[\protect\citeauthoryear{Sung et al.}{1996}]{saa}
Sung, H., Lee, S., Lee, M. G. \& Ann, H. B., 1996, J. Kor. Astr. So., 29, 269

\bibitem[\protect\citeauthoryear{Tadross}{2009}]{ta09}
Tadross, A. L., 2009, AP{\&}SS, 323 Issue-4, 383 

\bibitem[\protect\citeauthoryear{Wright et al.}{2010}]{wri}
Wright, E. L., Eisenhardt, P. R. M., Mainzer, A. K., et al. 2010, AJ, 140, 1868

\bibitem[\protect\citeauthoryear{Yadav et al.}{2013}]{yad}
Yadav, R. K. S., Sariya, D. P., \& Sagar R., 2013, MNRAS, 430, 3350 

\bibitem[\protect\citeauthoryear{Zacharias et al.}{2013}]{zac}
Zacharias, N., Finch, C. T., Girard, T. M., Henden, A., Bartlett, J. L., Monet, D. G. \& Zacharias, M. I., 2013, AJ, 145, 44

\bibitem[\protect\citeauthoryear{Zhao {\&} Shao}{1994}]{zha} 
 Zhao, J. L. \& Shao, Z. Y. 1994, A{\&}A, 288, 89

\end{thebibliography}

\end{document}